\documentclass[aps,prd,reprint,groupedaddress]{revtex4-1}
\usepackage{amsmath,amsfonts,amssymb,bm}
\usepackage{graphicx}
\usepackage{color}
\usepackage{subfigure}
\usepackage{multirow}
\usepackage{textcomp}
\usepackage{slashed}
\usepackage{cancel}

\definecolor{purple}{rgb}{0.5,0,0.5}
\definecolor{blue}{rgb}{0.0,0,0.9}

\usepackage[colorlinks=true, pdfstartview=FitV, linkcolor=purple, citecolor= purple, urlcolor=blue]{hyperref}

\begin{document}


\title{A Pattern for the Flavor Dependent Quark-antiquark Interaction}




\author{Muyang Chen}
\affiliation{School of Physics, Nankai University, Tianjin 300071, China}

\author{Lei Chang}\email{leichang@nankai.edu.cn}
\affiliation{School of Physics, Nankai University, Tianjin 300071, China}

\date{\today}

\begin{abstract}
A flavor dependent kernel is constructed based on the rainbow-ladder truncation of the Dyson-Schwinger and Bethe-Salpeter equation approach of Quantum Chromodynamics.
 The quark-antiquark interaction is composed of a flavor dependent infrared part and a flavor independent ultraviolet part.
 Our model gives a successful and unified description of the light, heavy and heavy-light ground pseudoscalar and vector mesons.
 For the first time, our model shows that the infrared enhanced quark-antiquark interaction is stronger and wider for the lighter quark.
\end{abstract}

\maketitle


\section{Introduction}
\label{sec:intro} 
\noindent

Hadrons, subatomic particles that are composed of quarks and gluons, perform a large scope of spectra:
the lightest hadron, the pion, has the mass $M_\pi \approx 0.14 \textmd{ GeV}$, while the heavy hadrons are heavier than $10 \textmd{ GeV}$\cite{Tanabashi2018}. 
People expect the underling theory, Quantum Chromodynamics(QCD)\cite{Brodsky1998}, can illuminate the hadron spectrum and unify the description of light and heavy hadrons.
QCD is a non-Abelian local gauge field theory of strong interaction and has been consistent with experimental observation up to present.
Due to the emergent phenomena at the hadronic scale, i.e. confinement and dynamical chiral symmetry breaking(DCSB), nonperturbative QCD is still the part of wilderness in the Standard Model.

Confinement provides an intrinsic wavelength, $\lambda_c\approx 0.5 \textmd{ fm}$, for the propagation of quarks and gluons.
They behave like the parton at $r<\lambda_c$ and show different propagation mode at $r>\lambda_c$. 
 The propagation of quarks and gluons would certainly be affected by the finite size of the hadrons.
Surveying hadron physics by QCD needs non-perturbative method. 
As an well-established non-perturbative approach, Lattice QCD(lQCD)\cite{Wilson1974,Ginsparg1982,Batrouni1985}, a lattice gauge theory formulated on a grid, has gained many progresses on the hadron physics.
While lQCD resorts to brute-force calculation, functional method like the Dyson-Schwinger equation and Bethe-Salpeter equation (DSBSE)\cite{Dyson1949,Schwinger1951,Roberts1994} approach is complementary to lQCD.


In this work, we aim at unifying the description of light, heavy and heavy-light mesons via DSBSE approach. 
In this Poincar$\acute{\text{e}}$ covariant framework the quark propagator is presented by the Gap equation\cite{Dyson1949,Schwinger1951,Roberts1994} 
\footnote{We work in the Euclidean space, where the inner product of the four vector is defined by $a\cdot b = \delta_{\mu\nu}a_\mu b_\nu = \sum_{i=1}^{4}a_i b_i$,
with $\delta_{\mu\nu}$ being the Kronecker delta. The Dirac matrices satisfy the algebra ${\gamma_\mu,\gamma_\nu} = 2\delta_{\mu\nu}$, and $\gamma_5=-\gamma_1\gamma_2\gamma_3\gamma_4$.},
\begin{eqnarray}\nonumber
 S_f^{-1}(k) &=& Z_2 (i\gamma\cdot k + Z_m m_f) \\\label{eq:DSE0}
&&+ \frac{4}{3}\bar{g}^2 Z_1 \int^\Lambda_{d q} D_{\mu\nu}(l)\gamma_\mu S_{f}(q)\Gamma^f_\nu(k,q),
\end{eqnarray}
where $f=\{u,d,s,c,b,t\}$ represents the quark flavor, $l=k-q$, $S_{f}$ the quark propagator, $m_f$ the current quark mass, $\Gamma^f_\nu$ the quark-gluon-vertex, $D_{\mu\nu}$ the gluon propagator, $\bar{g}$ the coupling constant.
$Z_1$, $Z_2$, $Z_m$ are the renormalisation constants of the quark-gluon-vertex, the quark field and the quark mass respectively.
$\int^\Lambda_{d q}=\int ^{\Lambda} d^{4} q/(2\pi)^{4}$ stands for a Poincar$\acute{\text{e}}$ invariant regularized integration, with $\Lambda$ the regularization scale.
A meson corresponds to a pole in the quark-antiquark scattering kernel\cite{Itzykson1980}, and the Bethe-Salpeter amplitude(BSA), $\Gamma^{fg}(k;P)$, with $k$ and $P$ the relative and the total momentum of the meson, $P^2 = -M^2$ and $M$ the meson mass, is solved by the Bethe-Salpeter equation(BSE)\cite{Itzykson1980,Salpeter1951,Roberts1994},
\begin{equation}\label{eq:BSE0}
  \big{[} \Gamma^{fg}(k;P)  \big{]}^{\alpha}_{\beta}  =   \int^\Lambda_{d q} \big{[} K^{fg}(k,q;P) \big{]}^{\alpha\delta}_{\sigma\beta} \big{[} \chi^{fg}(q;P)  \big{]}^{\sigma}_{\delta} ,
\end{equation}
where $ K^{fg}(k,q;P)$ is the quark-antiquark scattering kernel, and $\alpha$, $\beta$, $\sigma$ and $\delta$ are the Dirac indexes.
$\chi^{fg}(q;P) = S_{f}(q_{+}) \Gamma^{fg}(q;P) S_{g}(q_{-})$ is the wave function,
$q_{+} = q + \iota P/2$, $q_{-} = q - (1-\iota) P/2$, $\iota$ is the partitioning parameter describing the momentum partition between the quark and antiquark and dosen't affect the physical observables.

A promised consistent way to solve the problem of the meson spectrum is building a quark-gluon-vertex and constructing a scattering kernel.
The forms of the quark-gluon-vertex and scattering kernel have been investigated\cite{Chang2009}, while the most widely used and technically simple one is the rainbow-ladder(RL) approximation,
\begin{eqnarray}\label{eq:quarkRL}
&& \bar{g}^2 Z_1 D_{\mu\nu}(l) \Gamma^f_\nu(k,q) \to [Z_{2}]^{2} \tilde{D}^{f}_{\mu\nu}(l) \gamma_\nu, \\\label{eq:mesonRL}
&& \big{[} K^{fg}(k,q;P) \big{]}^{\alpha\delta}_{\sigma\beta} \to -\frac{4}{3}[Z_{2}]^{2} \tilde{D}^{fg}_{\mu\nu}(l) [\gamma_{\mu}^{}]^{\alpha}_\sigma [\gamma_{\nu}]^\delta_\beta,
\end{eqnarray}
where $\tilde{D}^{fg}_{\mu\nu}(l) = \left(\delta_{\mu\nu}-\frac{l_{\mu}l_{\nu}}{l^{2}}\right)\mathcal{G}^{fg}(l^2)$ and $\tilde{D}^{f}_{\mu\nu}(l) = \left(\delta_{\mu\nu}-\frac{l_{\mu}l_{\nu}}{l^{2}}\right)\mathcal{G}^f(l^2)$ are the effective quark-antiquark interaction.
In the original RL approximation, $\mathcal{G}^{fg} = \mathcal{G}^{f}$ is flavor symmetrical and modeled by\cite{Qin2011}
\begin{eqnarray}\label{eq:gluonmodel}
  \mathcal{G}^f(s) 	&=& \mathcal{G}^f_{IR}(s) + \mathcal{G}_{UV}(s),\\\label{eq:gluonInfrared}
  \mathcal{G}^f_{IR}(s) &=& 8\pi^2\frac{D_f^2}{\omega_f^4} e^{-s/\omega_f^2},\\\label{eq:gluonUltraviolet}
  \mathcal{G}_{UV}(s) 	&=& \frac{8\pi^{2} \gamma_{m}^{} \mathcal{F}(s)}{\text{ln}[\tau+(1+s/\Lambda^{2}_{QCD})^2]},
\end{eqnarray}
where $s=l^2$. $\mathcal{G}^f_{IR}(s)$ is the infrared interaction responsible for DCSB, with $D_f^2\omega_f$ expressing the interaction strength and $\omega_f$ the interaction width in the momentum space.
The form, Eq.(\ref{eq:gluonInfrared}), is used as it enables the natural extraction of a monotonic running-coupling and gluon mass\cite{Qin2011}, whose relation to QCD could be traced\cite{Binosi2015}.
$\mathcal{G}_{UV}(s)$ keeps the one-loop perturbative QCD limit in the ultraviolet.
$\mathcal{F}(s)=[1 - \exp(-s^2/[4m_{t}^{4}])]/s$, $\gamma_{m}^{}=12/(33-2N_{f})$, with $m_{t}=1.0 \textmd{ GeV}\,$, $\tau=e^{10} - 1$, $N_f=5$, and $\Lambda_{\text{QCD}}=0.21 \textmd{ GeV}\,$.
The values of $m_{t}$ and $\tau$ are chosen different from Ref.\cite{Qin2011} so that $\mathcal{G}_{UV}(s)$ is well suppressed in the infrared and the dressed function $\mathcal{G}_{IR}^{fg}(s)$ is qualitatively right in the limit $m_f \to \infty$ or $m_g \to \infty$..

A nontrivial property of $\Gamma^f_\nu$ is its dependence on the quark flavor due to the dressing effect.
By the same token, $K^{fg}$ depends on the flavors of the scattering quark and antiquark.
For a unified description of the system with different quarks, the flavor dependence of $\Gamma^f_\nu$ and $K^{fg}$ should be taken into account properly whatever model forms are used.
The RL approximation is phenomenologically successful for the pseudoscalar and vector mesons\cite{Maris1997,Maris1998,Maris1999,Qin2011}.
The best parameters are $(D_f^2\omega_f)^{1/3} = 0.8\textmd{ GeV}$ and $\omega_f = 0.5\textmd{ GeV}$ for the light mesons\cite{Qin2011}, 
and $(D_f^2\omega_f)^{1/3} \approx [0.6,0.7] \textmd{ GeV}$ and $\omega_f = 0.8\textmd{ GeV}$ for the heavy mesons\cite{Chen2017,Qin2018}.
The strength decreases and $\omega_f$ increases as the quark mass increases, showing the fact that heavy-flavor quarks probe shorter distances than light-flavor quarks at the corresponding quark-gluon-vertexes\cite{Serna2017}.
The RL approximation fails to describe the heavy-light mesons due to the lack of flavor asymmetry in Eq.(\ref{eq:gluonmodel})-Eq.(\ref{eq:gluonUltraviolet}).
The spectrum has a larger error than the quarkonia and the decay constants are totally false\cite{Nguyen2011,Hilger2017}.
The heavy-light mesons problem has been surveyed for 20 years in this approach\cite{Ivanov1999,Roberts2004,Ivanov2007,Gomez-Rocha2015a,Gomez-Rocha2016,Binosi2019}, yet no satisfied solution was gained.

\section{Our model}
\label{sec:method}
\noindent

To introduce the flavor asymmetry, one should concern the axial-vector Ward-Takahashi identity(av-WTI), which guarantees the ground state pseudoscalar mesons as Goldstone bosons of DCSB\cite{Maris1997,Maris1998},
 \begin{eqnarray}\nonumber
 P_\mu \Gamma^{fg}_{5\mu}(k;P) &=& S^{-1}_f(k_+)i\gamma_5 + i\gamma_5 S^{-1}_g(k_-)\\\label{eq:WTI}
				&&- i[m_f + m_g]\Gamma^{fg}_5(k;P), 
\end{eqnarray}
where $\Gamma^{fg}_{5\mu}$ and $\Gamma^{fg}_5$ are the axial-vector and pseudoscalar vertex respectively.
Considering the equations of $S^{f,g}$, $\Gamma^{fg}_{5\mu}$ and $\Gamma^{fg}_5$ in the RL approximation, Eq.(\ref{eq:WTI}) leads to 
\begin{eqnarray}\nonumber
&& \int^\Lambda_{d q} \mathcal{G}^{fg}(s)\gamma_\alpha [S_{f}(q_+)i\gamma_5 + i\gamma_5 S_{g}(q_-)]\gamma_\beta = \\\label{eq:WTIRL}
&& \int^\Lambda_{d q} \gamma_\alpha [\mathcal{G}^{f}(s) S_{f}(q_+)i\gamma_5 + \mathcal{G}^{g}(s) i\gamma_5S_{g}(q_-)]\gamma_\beta.
\end{eqnarray}

Eq.(\ref{eq:WTIRL}) tells us that $\mathcal{G}^{fg}(s)$ is some medium value of $\mathcal{G}^{f}(s)$ and $\mathcal{G}^{g}(s)$.
Considering the scalar part of the propagator, $S^f(q) = -i\slashed{q}\sigma^f_v(q^2) + \sigma^f_s(q^2)$, we get $\mathcal{G}^{fg}(s) = (\sigma^f_s(q^2_+)\mathcal{G}^{f}(s) + \sigma^g_s(q^2_-)\mathcal{G}^{g}(s))/(\sigma^f_s(q^2_+) + \sigma^g_s(q^2_-))$.
It is well known that the infrared value of $\sigma^f_s(q^2)$ is proportion to the inverse of the interaction strength and the width of $\sigma^f_s(q^2)$ is proportion to $\omega_f$. 
We thus assume $\mathcal{G}^{fg}(s)$ to be
\begin{eqnarray}\label{eq:gluonfmodel}
  \mathcal{G}^{fg}(s) 	&=& \mathcal{G}^{fg}_{IR}(s) + \mathcal{G}_{UV}(s),\\\label{eq:gluonfInfrared}
  \mathcal{G}^{fg}_{IR}(s) 	&=& 8\pi^2\frac{D_f}{\omega_f^2}\frac{D_g}{\omega_g^2} e^{-s/(\omega_f\omega_g)}.
\end{eqnarray}
$\mathcal{G}_{UV}(s)$ is unchanged from Eq.(\ref{eq:gluonUltraviolet}), and as we are dealing with 5 active quarks, $\mathcal{G}_{UV}(s)$ is independent of the quark flavors.
The effective interation $\tilde{D}^{fg}_{\mu\nu}$ represents the total dressing effect of the gluon propagator and the two quark-gluon-vertexes.
Eq.(\ref{eq:gluonfInfrared}) means the quark and antiquark contribute equally to the interaction strength and width.

\begin{figure}
 \includegraphics[width=\columnwidth]{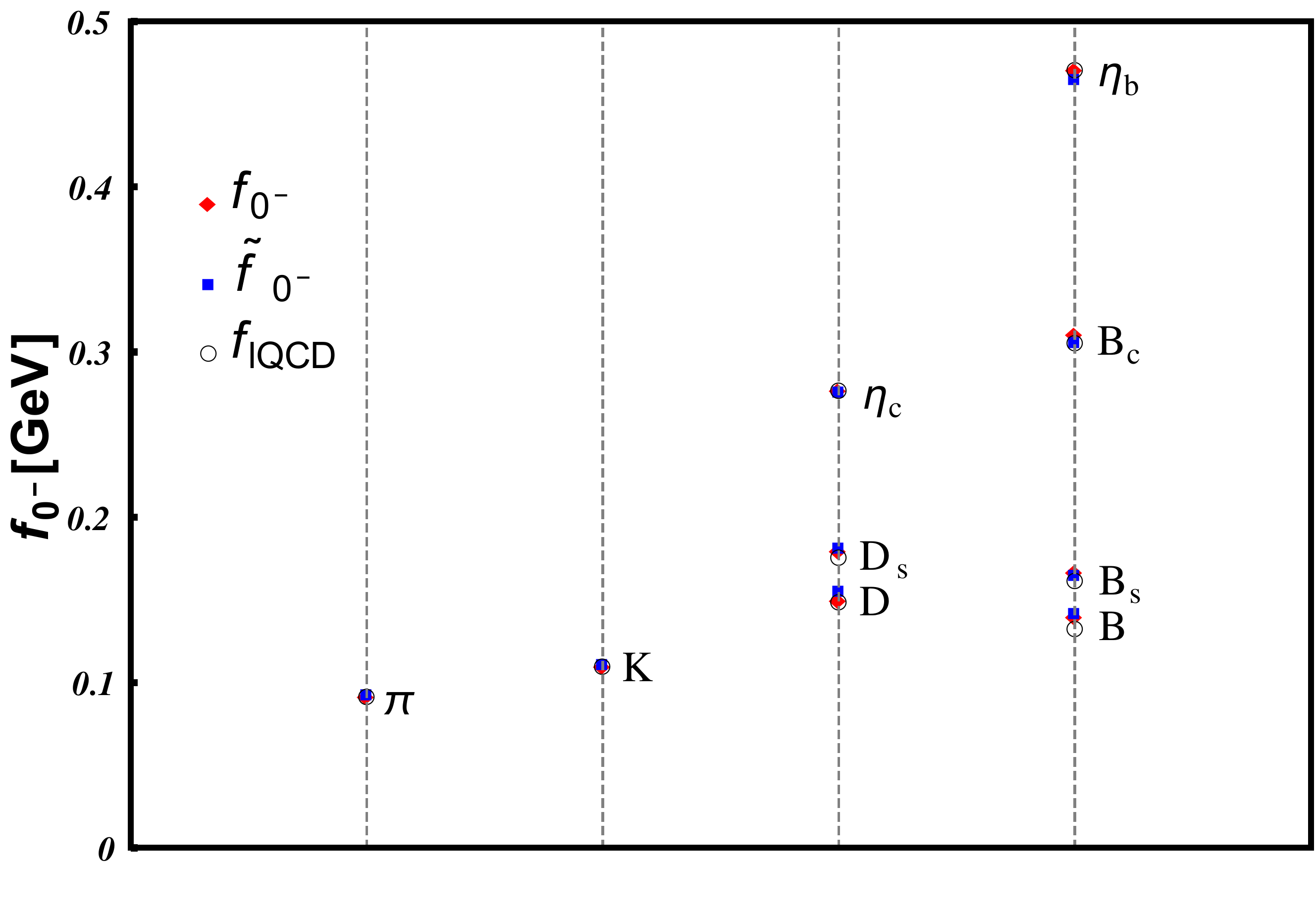}
\caption{\label{fit:decayconstants} Decay constants of the ground state pseudoscalar mesons: $f_{0^-}$ defined by Eq.(\ref{eq:definition-f}) and 
$\tilde{f}_{0^-}$ defined by Eq.(\ref{eq:GMOR}) and Eq.(\ref{eq:rho}) are our results, $f_{\textmd{lQCD}}$ are the lattice QCD data in Tab. \ref{tab:mesons0-}.}
\end{figure}

The preservation the av-WTI could be checked numerically by the Gell-Mann--Oakes--Renner(GMOR) relation, which is equivalent to the av-WTI\cite{Maris1997,Maris1998},
\begin{equation}\label{eq:GMOR}
 \tilde{f}_{0^-} := (m_f + m_g)\rho_{0^-}/M^2_{0^-} = f_{0^-},
\end{equation}
with $M_{0^-}$ the mass of the pseudoscalar meson and $f_{0^-}$ the leptonic decay constant. $f_{0^-}$ and $\rho_{0^-}$ are defined by
\begin{eqnarray}\label{eq:definition-f}
 f_{0^{-}}P_{\mu} &:=& Z_{2} N_{c} \;\text{tr} \! \int^{\Lambda}_{d k} \! \gamma_{5}^{} \gamma_{\mu}^{} S_f(k_+)\Gamma^{fg}_{0^{-}}(k;P)S_g(k_-),\\\label{eq:rho}
 \rho_{0^{-}} &:=& Z_{4} N_{c} \;\text{tr} \! \int^{\Lambda}_{d k} \! \gamma_{5}^{} S_f(k_+)\Gamma^{fg}_{0^{-}}(k;P)S_g(k_-),
\end{eqnarray}
with  $Z_4 = Z_2 Z_m$, $N_c$ the color number, $\text{tr}$ the trace of the Dirac index and $\Gamma^{fg}_{0^{-}}$ the BSA of pseudoscalar mesons.
The BSA is normalized by \cite{Nakanishi1965}
\begin{eqnarray}\nonumber
 2P_\mu &=& N_c \frac{\partial}{\partial P_\mu}\int^{\Lambda}_{d q} \textmd{tr} [ \Gamma(q;-K) \\\label{eq:normalization}
 &&\times S(q_+)\Gamma(q;K)S(q_-) ]|_{P^2=K^2=-M^2},
\end{eqnarray}
where $N_c=3$ is the color number.
Before discussing the details and results, we first assure the reader of the preservation of the av-WTI by comparing $f_{0^-}$ and $\tilde{f}_{0^-}$ in Fig.\ref{fit:decayconstants}.
They deviate from each other by no more than $3\%$ for all the pseudoscalar mesons considered here.
We conclude that the av-WTI is perfectly preserved in our approach.


\section{Outputs of the model}
\noindent

In Eq.(\ref{eq:gluonfInfrared}), $D_{f,g}$ and $\omega_{f,g}$ are parameters expressing the flavor dependent quark-antiquark interaction.
However, the flavor dependence of these parameters is apriori unknown.
Herein we treat the $D_{f}$ and $\omega_{f}$ of each flavor as free parameters.
Working in the isospin symmetry limit, we have 4 independent quarks up to the $b$ quark mass: $u$ (or $d$), $s$, $c$ and $b$.
There are 3 parameters for each flavor: $D_f$, $\omega_f$ and $m_f$.
In total there are 12 parameters.
$\omega_u$ is treated as an independent variable, the other 11 parameters are dependent variables, which are fitted by 11 observables: the masses and decay constants of $\pi$, $K$, $\eta_c$ and $\eta_b$, 
and the masses of $D$, $D_s$ and $B$.
All the masses and decay constants of the rest ground state pseudoscalar mesons(except $\eta$ and $\eta^\prime$) and all the ground state vector mesons are predicted.

The masses and decay constants of the ground state pseudoscalar mesons are listed Tab.\ref{tab:mesons0-}.
Our outputs are quite stable with $\omega_u$ varying $10\%$ around $0.5 \textmd{ GeV}$. 
With $\omega_u \in [0.45,0.55] \textmd{ GeV}$, the masses are almost unchanged and the decay constants vary within $1.2\%$.
Our output of $M_{B^{\pm}_s}$ deviates from the experimental value by only $0.01 \textmd{ GeV}$, which is impossible in the original RL truncated DSBSE.
The flavor dependence of the quark gluon interaction even has a significant effect on the $B_c$ meson.
$M_{B_c}$ produced by the original RL truncated DSBSE is $0.11\textmd{ GeV}$ larger than the experimental value\cite{Qin2018}.
We reduce the error to less than $0.02 \textmd{ GeV}$ herein.
Our output of $f_D$, $f_{D^{\pm}_s}$, $f_{B}$, $f_{B^{\pm}_s}$ and $f_{B_c}$ are all consistent with the lattice QCD results, with the deviation less than $6\%$.
Note that our $f_{D^{\pm}_s}$ is also in good agreement with the recent experimental measurement\cite{Ablikim2019}.
The only absent mesons in Tab.\ref{tab:mesons0-} are $\eta$ and $\eta^\prime$, which are affected by the axial anomaly\cite{Bhagwat2007,Ding2019} and beyond our present purpose.

\begin{table}
\caption{\label{tab:mesons0-} Masses and decay constants of the ground state pseudoscalar mesons (in GeV). We use the convention $f_\pi = 0.093\textmd{ GeV}$.
The lQCD data are taken from: $M_D$ and $M_{D_s}$ - Ref. \cite{Cichy2016}; $M_B$ and $M_{B_s}$ - Ref. \cite{Dowdall2012};
$M_{B_c}$ - Ref. \cite{Mathur2018}; $f_\pi$ and $f_K$ - Ref. \cite{Follana2008}; $f_D$, $f_{D_s}$, $f_B$ and  $f_{B_s}$ - Ref. \cite{Bazavov2018};
$f_{\eta_c}$ and $f_{\eta_b}$ - Ref. \cite{McNeile2012}; $f_{B_c}$ - Ref. \cite{Colquhoun2015}.
$M_{\pi}$, $M_K$, $M_{\eta_c}$, $M_{\eta_b}$ here and $M_\Upsilon$ in Tab.\ref{tab:mesons1-} are usually used to fit the quark masses in lQCD calculations\cite{Aoki2017},
so there are no lQCD predictions for these quantities.
The exprimental data are taken from Ref. \cite{Tanabashi2018}.
Note that we work in the isospin symmetry limit, so the average value among or between the isospin multiplet is cited for $\pi$, $K$, $D$ and $B$ meson.
All the date are cited with accuracy $0.001 \textmd{ GeV}$.
In our production, the underlined values are those used to fit the 11 dependent variables, 
and the others are our output with the uncertainty corresponding to $\omega_u \in [0.45,0.55] \textmd{ GeV}$.
The decay constants are fitted to the lQCD data because an accurate and complete experimental estimate of these data is still lacking.
}
\begin{tabular}{c|c|c|c|c|c|c|c}
\hline
		&	herein		&	lQCD	&	expt.	&\hspace*{-0.1cm}&	\hspace*{-0.1cm}	&	herein		&	lQCD		\\
\cline{1-4}\cline{6-8}
$M_{\pi}$ 	&\underline{0.138}	&	$\ast$	&\underline{0.138(1)}	&\hspace*{-0.1cm}&$f_{\pi}$		&\underline{0.0093}	&\underline{0.0093(1)}		\\
$M_K$	    	&\underline{0.496}	&	$\ast$	&\underline{0.496(1)}	&\hspace*{-0.1cm}&$f_K$			&\underline{0.111}	&\underline{0.111(1)}		\\
$M_D$	    	&\underline{1.867}	&	1.865(3)&\underline{1.867(1)}	&\hspace*{-0.1cm}&$f_D$			&0.151(1)		&0.150(1)			\\
$M_{D^{\pm}_s}$	&\underline{1.968}	&	1.968(3)&\underline{1.968(1)}	&\hspace*{-0.1cm}&$f_{D^{\pm}_s}$	&0.181(1)		&0.177(1)			\\
$M_{\eta_c}$	&\underline{2.984}	&	$\ast$	&\underline{2.984(1)}	&\hspace*{-0.1cm}&$f_{\eta_c}$		&\underline{0.278}	&\underline{0.278(2)}		\\
$M_{B}$		&\underline{5.279}	&	5.283(8)&\underline{5.279(1)}	&\hspace*{-0.1cm}&$f_{B}$		&0.141(2)		&0.134(1)			\\
$M_{B^{\pm}_s}$	&5.377(1)		&	5.366(8)&5.367(1)		&\hspace*{-0.1cm}&$f_{B^{\pm}_s}$	&0.168(2)		&0.163(1)			\\
$M_{B_c}$	&6.290(3)		&	6.276(7)&6.275(1)		&\hspace*{-0.1cm}&$f_{B_c}$		&0.312(1)		&0.307(10)			\\
$M_{\eta_b}$	&\underline{9.399}	&	$\ast$	&\underline{9.399(2)}	&\hspace*{-0.1cm}&$f_{\eta_b}$		&\underline{0.472}	&\underline{0.472(5)}		\\  
\hline
\end{tabular}
\end{table}

\begin{table}
\caption{\label{tab:mesons1-} Masses and decay constants of ground state vector mesons (in GeV).
The lQCD data are taken from: $M_{\rho}$ - Ref. \cite{Fu2016}; $M_{K^*}$ - Ref. \cite{Dudek2014}; $M_\phi$ and $f_\phi$ - Ref. \cite{Donald2014a};
$M_{D^*}$, $f_{D^*}$, $M_{D^{*\pm}_s}$, $f_{D^{*\pm}_s}$, $M_{B^*}$, $f_{B^*}$, $M_{B^{*\pm}_s}$ and $f_{B^{*\pm}_s}$ - Ref. \cite{Lubicz2017};
$M_{J/\psi}$ and $f_{J/\psi}$ - Ref. \cite{Donald2012};
$M_{B^*_c}$ - Ref. \cite{Mathur2018}; $f_{B^*_c}$ - Ref. \cite{Colquhoun2015};
$f_\Upsilon$ - Ref. \cite{Colquhoun2015a}.
The exprimental data are taken from Ref. \cite{Tanabashi2018}, the average value between the isospin multiplet is cited for $M_{D^*}$.
Hitherto, $B^*_c$ meson is not discovered experimentally.
All the date are cited with accuracy $0.001 \textmd{ GeV}$.
The uncertainties of our results correspond to $\omega_u \in [0.45,0.55] \textmd{ GeV}$.
}
\begin{tabular}{c|c|c|c|c|c|c|c}
\hline
		&herein		&lQCD		&expt.		&\hspace*{-0.1cm}&			&herein		&lQCD		\\
\cline{1-4}\cline{6-8}
   $M_{\rho}$	&0.724(2)	&0.780(16)	&0.775(1)	&\hspace*{-0.1cm}&$f_{\rho}$		&0.149(1)	&--		\\  
   $M_{K^*}$	&0.924(2)	&0.933(1)	&0.896(1)	&\hspace*{-0.1cm}&$f_{K^*}$		&0.160(2)	&--		\\  
   $M_\phi$	&1.070(1)	&1.032(16)	&1.019(1)	&\hspace*{-0.1cm}&$f_\phi$		&0.191(1)	&0.170(13)	\\  
   $M_{D^*}$	&2.108(4)	&2.013(14)	&2.009(1)	&\hspace*{-0.1cm}&$f_{D^*}$		&0.174(4)	&0.158(6)	\\  
$M_{D^{*\pm}_s}$&2.166(7)	&2.116(11)	&2.112(1)	&\hspace*{-0.1cm}&$f_{D^{*\pm}_s}$	&0.206(2)	&0.190(5)	\\  
   $M_{J/\psi}$	&3.132(2)	&3.098(3)	&3.097(1)	&\hspace*{-0.1cm}&$f_{J/\psi}$ 		&0.304(1)	&0.286(4)	\\  
   $M_{B^*}$	&5.369(5)	&5.321(8)	&5.325(1)	&\hspace*{-0.1cm}&$f_{B^*}$		&0.132(3)	&0.131(5)	\\  
$M_{B^{*\pm}_s}$&5.440(1)	&5.411(5)	&5.415(2)	&\hspace*{-0.1cm}&$f_{B^{*\pm}_s}$	&0.152(2)	&0.158(4)	\\  
   $M_{B^*_c}$	&6.357(3)	&6.331(7)	&--		&\hspace*{-0.1cm}&$f_{B^*_c}$		&0.305(5)	&0.298(9)	\\  
   $M_\Upsilon$	&9.454(1)	&$\ast$		& 9.460(1)	&\hspace*{-0.1cm}&$f_\Upsilon$		&0.442(3)	&0.459(22)	\\  
\hline
\end{tabular}
\end{table}

A further confirmation of our model is given by the vector mesons.
Our predictions of the static vector meson masses and decay constants are listed in Tab.\ref{tab:mesons1-}.
The decay constant is defined analogy to Eq.(\ref{eq:definition-f})
\begin{equation}
 f_{1^{-}}M_{1^-} = Z_{2} N_{c} \;\text{tr} \! \int^{\Lambda}_{d k} \! \gamma_{\mu}^{} S_f(k_+)\Gamma^{\mu,fg}_{1^{-}}(k;P)S_g(k_-),
\end{equation}
with $M_{1^-}$ the vector meson and $\Gamma^{\mu,fg}_{1^{-}}$ the vector meson BSA.
The vector mesons also show a weak dependence on $\omega_u \in [0.45,0.55] \textmd{ GeV}$.
The deviation from experimental or lQCD values decreases as the mass increases.
The mass deviation is about $6\%$ for the $\rho$ meson, decreasing to about $1\%$ for the heavy mesons.
The decay constant deviation is about $12\%$ for the $\phi$ meson, decreasing to less than $7\%$ for the heavy mesons.
This deviation is attributed to the systematic error of the RL truncation\cite{Maris1999}.
The successfulness of the pattern of the flavor dependent interaction, Eq.(\ref{eq:gluonfmodel},\ref{eq:gluonfInfrared},\ref{eq:gluonUltraviolet}), is shown by the fact that the deviation is of the same order for both the open-flavor mesons and the quarkonia.
We can see again that the flavor dependence has a significant effect on $B_c$ mesons.
While $M_{B^*_c}\approx 6.54 \textmd{ GeV}$ and $f_{B^*_c} \approx 0.43 \textmd{ GeV}$ in the original RL truncated DSBSE\cite{Qin2018},
our results $M_{B^*_c} \approx 6.357 \textmd{ GeV}$ and $f_{B^*_c} \approx 0.305 \textmd{ GeV}$ is more consistent with the lQCD predictions.
$B^*_c$ has not been discovered experimentally, both our and lQCD predictions wait for the experimental verification.

\begin{figure}\hspace*{0.25em}
 \includegraphics[width=0.439\textwidth]{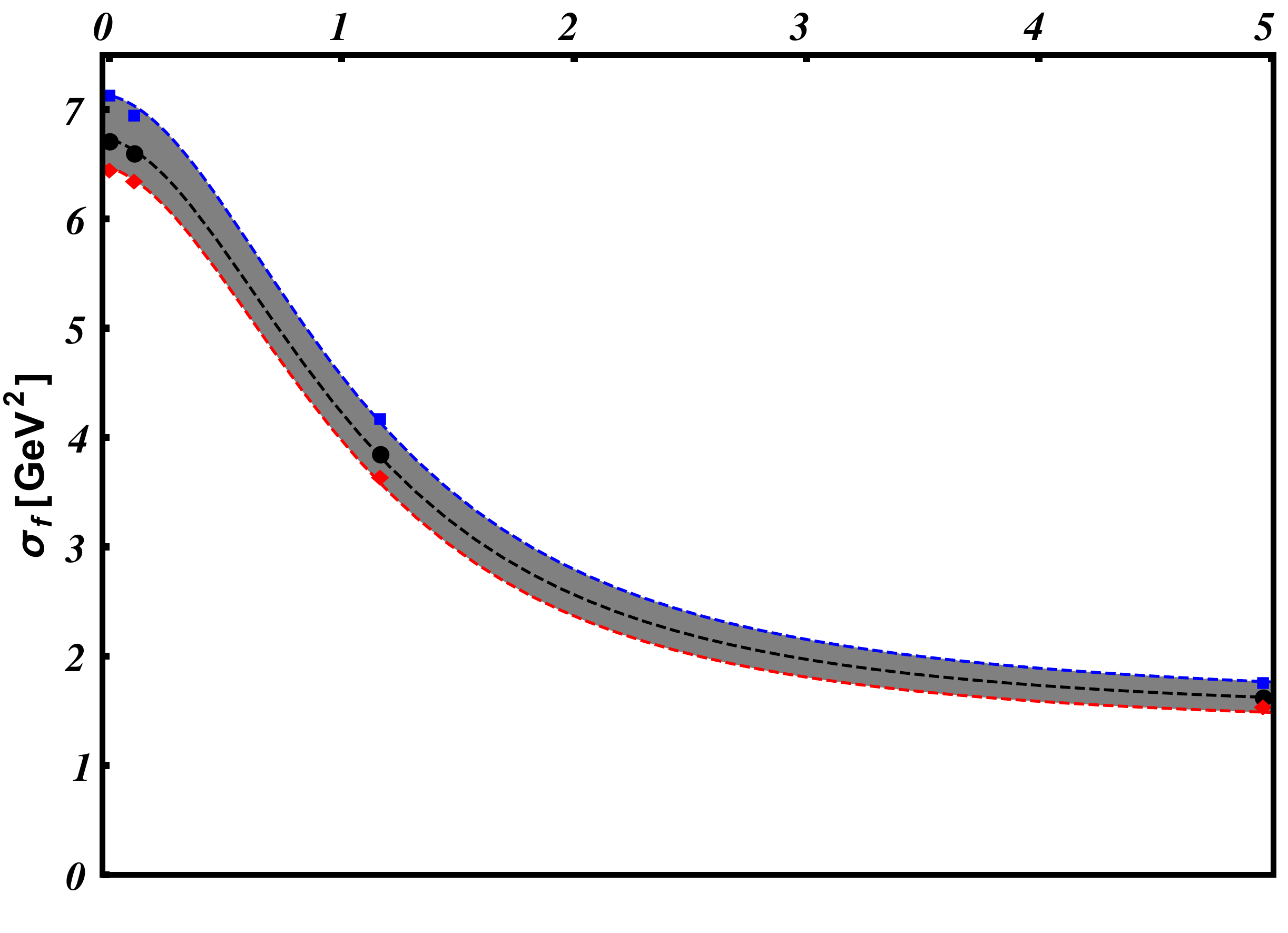}\vspace*{-1.0em}
 \includegraphics[width=0.45\textwidth]{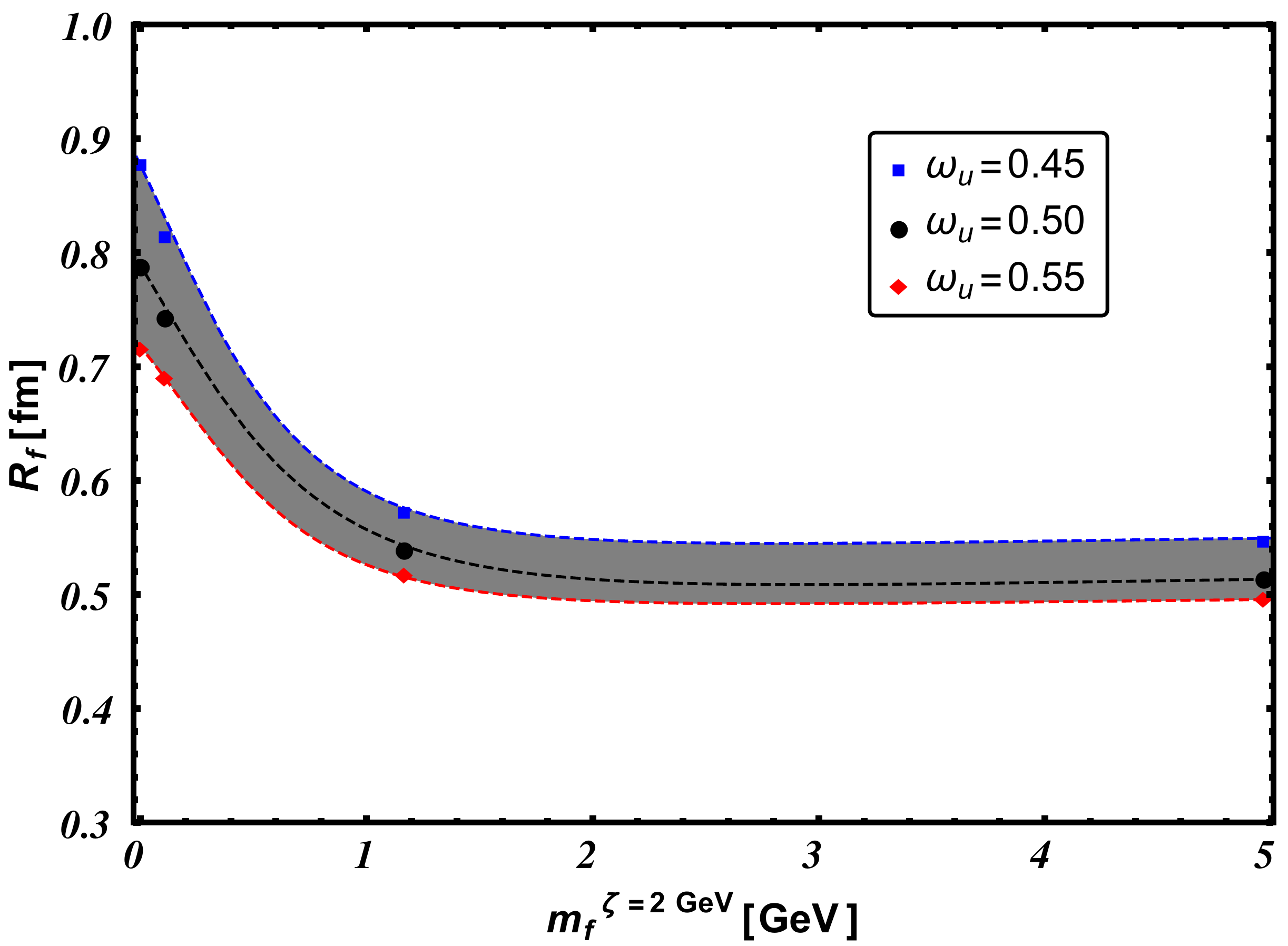}
\caption{\label{fig:wf} Quark mass dependence of the interaction tension $\sigma_f$ and the radius $R_{f}$.
The lines are drawn to guide eyes.}
\end{figure}

At last, we investigate the flavor dependence of the quark-antiquark interaction.
In the heavy quark limit, the dressing of the quark-gluon-vertex might be ignored and our adopted interaction is in agreement with QCD\cite{Binosi2015}, 
so the interaction should saturate $\mathcal{G}^{ff}(l^2) \xrightarrow{m_f\to\infty} 4\pi\alpha_s \frac{Z(l^2)}{l^2}$,
with $\alpha_s$ the strong-interaction constant and $Z(l^2)$ the dressing function of the gluon propagator, defined by $\Delta_{\mu\nu}(l) = (\delta_{\mu\nu}-\frac{l_\mu l_\nu}{l^2})\frac{Z(l^2)}{l^2}$, with $\Delta_{\mu\nu}(l)$ being the dressed gluon propagator.
As we fix $N_f=5$, both $\alpha_s$ and  $Z(l^2)$ are independent of the interacting quarks.
Phenomenally the parameters, $D_f$ and $\omega_f$, should go to a constant as the quark mass increases.
In the chiral limit the interaction is enhanced because of the dressing of the quark-gluon-vertex\cite{Alkofer2009,Williams2015,Binosi2017,Sternbeck2017,Oliveira2018}, which is necessary to trigger chiral symmetry breaking.
The potential is properly defined by the fourier-transform of the interaction. 
For the interesting infrared part of our model we have $\mathcal{V}_{\textmd{IR}}^{ff}(\vec{r}) = \int \textmd{d}^3\vec{l}\, \mathcal{G}_{\textmd{IR}}^{ff}(l^2)\textmd{e}^{-\vec{l}\cdot\vec{r}/\omega^2_f} \propto \text{e}^{-\vec{r}^2/R_{f}^2}$, with $\vec{r}$ the space coordinate and $R_{f}=2/\omega_f$ expressing the radius of the quark-gluon interaction.
Additionally, we adopt the following quantity to describe the interaction strength\cite{Roberts2000}:
\begin{equation}
 \sigma_f = \frac{1}{4\pi}\int_{\Lambda_{\textmd{QCD}}^2}^{(10\Lambda_{\textmd{QCD}})^2}\textmd{ds}\,\mathcal{G}^{ff}(s)*s.
\end{equation}
The quark mass dependence of $\sigma_f$ and $R_{f}$ is depicted in Fig.\ref{fig:wf}.
The interaction strength and radius reduce as quark mass raises, which is expected by the fact that the quark-gluon-vertex dressing effect should decrease as quark mass increases\cite{Williams2015}.
The interaction radius, $2/\sqrt{\omega_{f}\omega_{g}}$, also expresses another fact that the quarks and gluons have a maximum wavelength of the hadron size\cite{Brodsky2008}.

\section{Summary and conclusion}
\label{sec:conclusion}
\vspace{0.1cm}

In summary, the flavor dependence of the full quark-antiquark interaction is an intrinsic property of QCD, and crucial for an unified description of light and heavy hadrons.
While a perfect quark-gluon-vertex that hold the proper flavor dependence of QCD has not been found, we construct a flavor dependent kernel based on the RL truncation of DSBSE.
The quark-antiquark interaction is composed of a flavor dependent infrared part and a flavor independent ultraviolet part.
With the parameters fixed by physical observables, our model takes into account not only the flavor dependence, but also the affection of the hadron size.
Our model, with the av-WTI perfectly preserved, provides a successful unified description of light, heavy and heavy-light ground pseudoscalar and vector mesons.
For the first time, our model shows that the infrared enhanced quark-antiquark interaction is stronger and wider for the lighter quark.
This flavor dependence pattern is universal, and is supposed to be applicable to baryons, for example, the double charm baryons within the Faddeev approach.
Our approach also provides a proper description of the inner structure of the heavy-light mesons, which can be used to calculate some scattering process, such as the $B$ to $\pi$ transition form factor.

\section*{Acknowledgments}

Work supported by: the Chinese Government’s Thousand Talents Plan for Young Professionals.

\appendix
\setcounter{figure}{0}
\setcounter{table}{0}
\renewcommand{\thefigure}{A\arabic{figure}}
\renewcommand{\thetable}{A\arabic{table}}

\section{Appendix A}\label{sec:appendixA}

The fitted parameters correspond to $\omega_u = 0.45, 0.50, 0.55 \textmd{ GeV}$ are listed in Tab.\ref{tab:parameters}.
The quark mass $\bar{m}_f^{\zeta}$ is defined by
\begin{eqnarray}
 \bar{m}_f^{\zeta} &=& \hat{m}_f/\left[\frac{1}{2}\textmd{Ln}\frac{\zeta^2}{\Lambda^2_{\textmd{QCD}}}\right]^{\gamma_m},\\
 \hat{m}_f	 &=& \lim_{p^2 \to \infty}\left[\frac{1}{2}\textmd{Ln}\frac{p^2}{\Lambda^2_{\textmd{QCD}}}\right]^{\gamma_m} M_f(p^2),
\end{eqnarray}
with $\zeta$ the renormalisation scale, $\hat{m}_f$ the renormalisation-group invariant current-quark mass and $M_f(p^2)$ the quark mass function, $S_f(p)=\frac{Z_f(p^2,\zeta^2)}{i\gamma\cdot p + M_f(p^2)}$.

\begin{table}[h!]
\caption{\label{tab:parameters} Fitted parameters correspond to $\omega_u = 0.45, 0.50, 0.55 \textmd{ GeV}$. $\bar{m}_f^{\zeta=2\textmd{GeV}}$, $\omega_f$ and $D_f$ are all measured in GeV.}
\begin{tabular}{c|c|c|c|c|c|c|c|c|c|c}
\hline \hline
flavor& $\bar{m}_f^{\zeta=2\textmd{GeV}}$&\hspace*{-0.1cm}&\; $w_f $\; &\; $D_f^2$ \;&\hspace*{-0.1cm}&\; $w_f $\; &\; $D_f^2$ \;	&\hspace*{-0.1cm}	&\; $w_f $\; &\; $D_f^2$ \;\\ [0.5mm]
\hline
$u$	& 0.0049&\hspace*{-0.1cm}	& 0.450 & 1.133 		&\hspace*{-0.1cm}	& 0.500 & 1.060 		&\hspace*{-0.1cm}	& 0.550 & 1.014 \\
$s$	& 0.112	&\hspace*{-0.1cm}	& 0.490 & 1.090 		&\hspace*{-0.1cm}	& 0.530 & 1.040 		&\hspace*{-0.1cm}	& 0.570 & 0.998 \\
$c$	& 1.17	&\hspace*{-0.1cm}	& 0.690 & 0.645 		&\hspace*{-0.1cm}	& 0.730 & 0.599 		&\hspace*{-0.1cm}	& 0.760 & 0.570 \\
$b$	& 4.97	&\hspace*{-0.1cm}	& 0.722 & 0.258 		&\hspace*{-0.1cm}	& 0.766 & 0.241 		&\hspace*{-0.1cm}	& 0.792 & 0.231 \\
\hline \hline
\end{tabular}
\end{table}

\bibliography{mesonsHL}

\end{document}